%% file: ICRC2025_template/skeleton.tex
\documentclass[a4paper,11pt]{article}
\usepackage{pos}
\usepackage{tabularray}
\usepackage[utf8]{inputenc}
\usepackage{textcomp}
\DeclareUnicodeCharacter{2212}{\ensuremath{-}}
\usepackage{wrapfig}
\usepackage{lineno}
\usepackage{enumitem}
\let\oldthebibliography\thebibliography
\let\endoldthebibliography\endthebibliography

\renewenvironment{thebibliography}[1]{%
  \oldthebibliography{#1}%
  \setlength{\itemsep}{0pt}%
  \setlength{\parskip}{0pt}%
  \setlength{\parsep}{0pt}%
}{%
  \endoldthebibliography
}

\title{Auger@TA: In-situ Cross-Calibration of the World's Largest Cosmic Ray Observatories}
\ShortTitle{Auger@TA}

\author*[a]{\small Adriel G B Mocellin}
\author{\small J. Cara\c{c}a-Valente}
\author{\small C. Covault}
\author{\small E. Dalcan}
\author{\small T. Fujii}
\author{\small S. Im}
\author{\small R. James}
\author{\small J.~Johnsen}
\author{\small K.H. Kampert}
\author{\small H. Kern}  
\author{\small J.N. Matthews}  
\author{\small E. Mayotte}  
\author{\small S. Mayotte}  
\author{\small X.~Moskala}
\author{\small H.~Que}
\author{\small J. Rautenberg}
\author{\small M. Roth}
\author{\small H. Sagawa}
\author{\small T. Sako}
\author{\small F. Sarazin}
\author{\small R. Sato}
\author{\small D. Schmidt}
\author{\small S.B.~Thomas}
\author{\small G. W\"orner}

\affiliation[a]{\footnotesize Department of Physics, Colorado School of Mines, Golden, CO, USA}
\affiliation[b]{\footnotesize Observatorio Pierre Auger, Av.\ San Mart{\'\i}n Norte 304, 5613 Malarg\"ue, Argentina
\\
Full author list: \normalfont{\url{https://www.auger.org/archive/authors\_icrc\_2023.html}}
}
\affiliation[c]{\footnotesize Telescope Array Project, 201 James Fletcher Bldg, 115 S 1400 East, Salt Lake City, UT 84112, USA\\

Full author list: \normalfont{\url{http://www.telescopearray.org/research/collaborators}}
}
\onbehalf{ for the Pierre Auger$^j$ and Telescope Array$^k$ Collaborations}
\emailAdd{ spokespersons@auger.org}
\onbehalf{for the Pierre Auger$^b$ and Telescope Array$^c$ Collaborations}

\abstract{The Pierre Auger Observatory (Auger) and the Telescope Array (TA) are the world’s two largest ultra-high-energy cosmic ray (UHECR) observatories. They operate in the Southern and Northern hemispheres, respectively, at similar latitudes but with distinct surface detector (SD) designs. A significant challenge in studying UHECR physics across the full sky is the apparent discrepancy in flux measurements between the two experiments. This discrepancy could arise from astrophysical differences and/or systematic effects related to their detector designs and sensitivities to extensive air shower components. To address this, the Auger@TA working group aims to cross-calibrate the two observatories with a self-triggering micro-Auger array within the TA array. This micro-array consists of eight Auger Surface Detector (SD) stations equipped with Water Cherenkov Detectors (WCDs) and AugerPrime Surface Scintillator Detectors. Seven SD stations, configured with a centered-1-PMT design, are arranged in a hexagonal pattern with one station in the center, with 1.5 km spacing, mirroring the Auger layout. The eighth station, which features a standard 3-PMT Auger station, is located in conjunction with a TA detector at the center of the hexagon, forming a triplet for high-statistics and low-uncertainty cross-calibration. A custom communication system that uses readily available components enables seamless communication between stations and remote access to each station through a central computer. The micro-array is now fully deployed, and initial data-taking is about to start. This presentation will detail the instrumentation, communication systems, central data acquisition system, expected performance of the micro-array, and preliminary results as appropriate.}

\ConferenceLogo{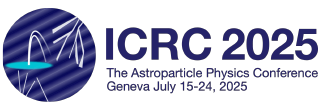}

\FullConference{39th International Cosmic Ray Conference (ICRC2025)\\
 15–24 July 2025\\
Geneva, Switzerland\\}

\begin{document}
\maketitle

\section{Introduction}

The Pierre Auger Observatory (Auger) and the Telescope Array (TA) are the two largest cosmic ray observatories in the world, covering areas of more than 3000 km$^2$ and 700 km$^2$, respectively. Situated in the southern and northern hemispheres, they provide complementary views of the sky, enabling extensive studies of ultra-high energy cosmic rays (UHECRs). Despite using different detection techniques and calibration methods, both observatories have independently observed a suppression in the cosmic ray flux at the highest energies \cite{verzi2017}.

Determining cosmic ray energies by surface detector arrays is essential for accurately characterizing the spectrum. However, discrepancies have been recorded in the energy spectra reported by the two experiments. To investigate these differences, the Auger and TA collaborations formed an experimental joint working group called Auger@TA to understand and reconcile the observed spectral variations.

Auger and TA employ different surface-detector reconstruction approaches to estimate primary energies, reflecting their detector geometries and calibration techniques. At Auger, the signal at 1000 m from the shower axis, S(1000), is converted to a reference zenith angle of 38$^{\circ}$ via the Constant Intensity Cut (CIC) method and then directly correlated with the fluorescence detector (FD) energy, producing an SD energy-scale statistical uncertainty of about 1\% at the highest energies \cite{auger20}.

\begin{figure}[h]
    \centering
    \includegraphics[width=0.8\textwidth]{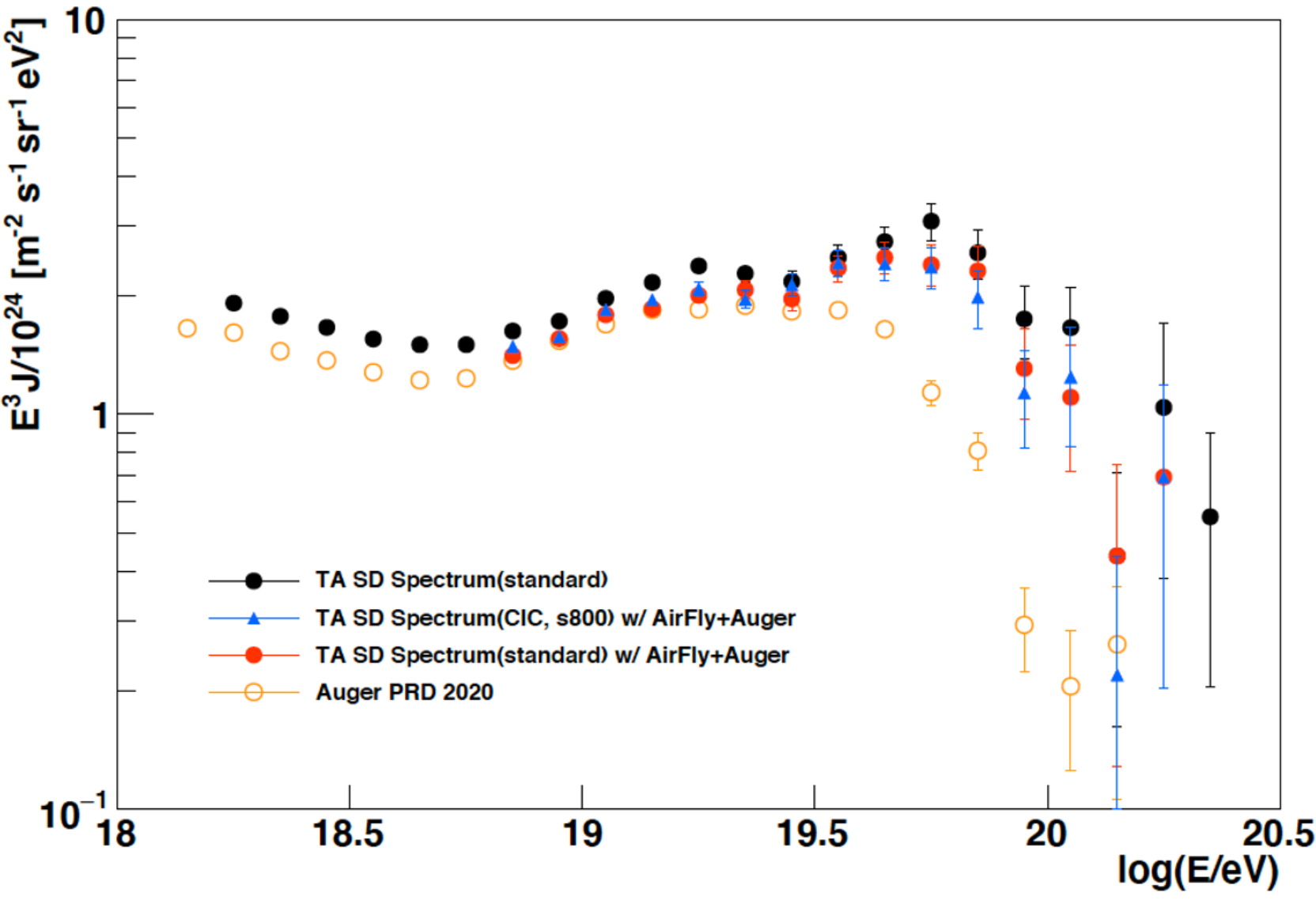}
    \caption{Comparison of the ultra-high energy cosmic-ray spectra measured by TA and Auger. Black points: TA standard analysis (14 years of SD data, TA fluorescence yield and missing‐energy corrections). Red points: TA with AirFly fluorescence yield and Auger missing‐energy correction (standard MC method). Blue triangles: TA with AirFly and Auger corrections using the CIC method. Orange open circles: Auger 2020 results \cite{kim_23}.}
    \label{fig:spectrum}
\end{figure}

TA, by contrast, derives its SD energy scale using Monte Carlo–derived lookup tables mapping the measured density at 800 m, S(800), to primary energy and zenith angle; an independent CIC-based reconstruction validated this simulation-based scale at about the 3\% level \cite{ivanov2019}. Since the Auger–TA Energy Spectrum Working Group convened in 2012 to cross-calibrate the two observatories, TA has investigated Auger’s fundamental analysis choices with the laboratory-measured AirFly model, which would lower TA energies by about 14\% when matched to Auger’s calibration \cite{pos_23}, switching to Auger’s experimentally derived missing-energy correction, and applying the CIC procedure to normalize S(800) to 38$^{\circ}$ exactly as Auger does for S(1000).

These combined alternatives appear to tighten the overall energy-scale offset from about 9\% to below 1\% for energies below $10^{19.5}$ eV (Figure 1), demonstrating that consistent fluorescence-yield and invisible-energy treatments may be adequate to bring the two spectra into near agreement, and motivating the Auger@TA project both to probe the remaining residual differences and to validate the robustness of these cooperative calibration procedures.

Moreover, studies report an $8\sigma$ significance in the variation of the UHECR energy spectrum between the northern and southern hemispheres, indicating possible differences between the northen and southern skies \cite{ta2024}. This observation raises some other critical questions: Are these spectral differences due to fundamental variations in the UHECR sky, or do they derive from unresolved systematic uncertainties between the two experiments? The Auger@TA micro-array is designed to probe whether local detector effects, or reconstruction alternatives can account for this residual difference.

This work presents the design, deployment, and early performance of a 7-station micro-array built to cross-calibrate the Auger and TA responses to the same ultra-high-energy cosmic ray showers. First, we describe a dedicated Central Data Acquisition System (CDAS) tailored to Auger@TA. Although it borrows many software components from the regular Auger CDAS, our version incorporates a trigger logic optimized for a smaller configuration. Second, we successfully commissioned one of the hexagon’s triangular segments, which is currently operating with four fully functional stations. Finally, we outline significant enhancements to the communications network—hardware and firmware—that ensure robust data transmission. We conclude with an outlook on completing the full hexagonal micro-array deployment, a comprehensive calibration strategy, and joint Auger–TA spectral analyses to resolve the remaining discrepancies in the UHECR spectrum.

\section{The Auger@TA Project}

The Auger@TA project was created with the goal of addressing the already cited discrepancies, focusing on a joint calibration effort between the two observatories. An Auger standard hexagonal micro-array, comprising seven stations equipped with single photomultiplier tubes (PMTs), was deployed within TA. At the center of this array, a triplet of stations was set for calibration purposes: a regular Auger station with a three-PMT setup, the Auger@TA station with one PMT, and an independent TA station, each being about 12 meters apart from one another. This setup aims to enable direct cross-calibration and to identify any underlying detector-based causes of the spectral discrepancies by comparing the same events with both detection approaches.

Both Auger and TA detect extensive air showers (EAS) using surface detector (SD) arrays, but each employs distinct mechanisms to collect particle signals. Auger uses Water Cherenkov Detectors (WCD), cylindrical tanks filled with ultra-purified water and equipped with PMTs. Charged particles from air showers generate Cherenkov radiation upon crossing through the water, producing a detectable light signal captured by the PMTs. Auger then calibrates the WCD signals using the Vertical Equivalent Muon (VEM), defined by the signal of a single vertical muon passing through the detector \cite{auger_wcd}. Additionally, Auger stations are also equipped with a Scintillator Surface Detector (SSD) installed on top of the WCD, providing complementary measurements and enhancing detector sensitivity.

In contrast, TA uses plastic scintillator detectors, where charged shower particles excite the plastic scintillator material, causing it to emit scintillation photons that are collected by wavelength-shifting fibers and guided to PMTs for detection \cite{ta_scintillator}. Their calibration employs the Minimum Ionizing Particle (MIP) as its calibration standard, corresponding to particles depositing the same amount of energy within the scintillator \cite{belz_2018}. To investigate and reconcile possible discrepancies, the intention is to simultaneously measure and reconstruct the same air showers with both detectors in situ. Auger@TA will enable detailed event-by-event cross-calibration, likely reducing systematic uncertainties, and facilitating a joint flux measurement.

% To support this effort, a dedicated and slightly modified CDAS is in its final stages of development. Currently, the first stations, including the center Auger doublet, are in operation, and allow for the data collection on a subset of the micro-array. Prospects include completing the hexagonal array, extending data collection, and starting a comprehensive analysis.

\begin{figure}[h]
    \centering
    \includegraphics[width=1\textwidth]{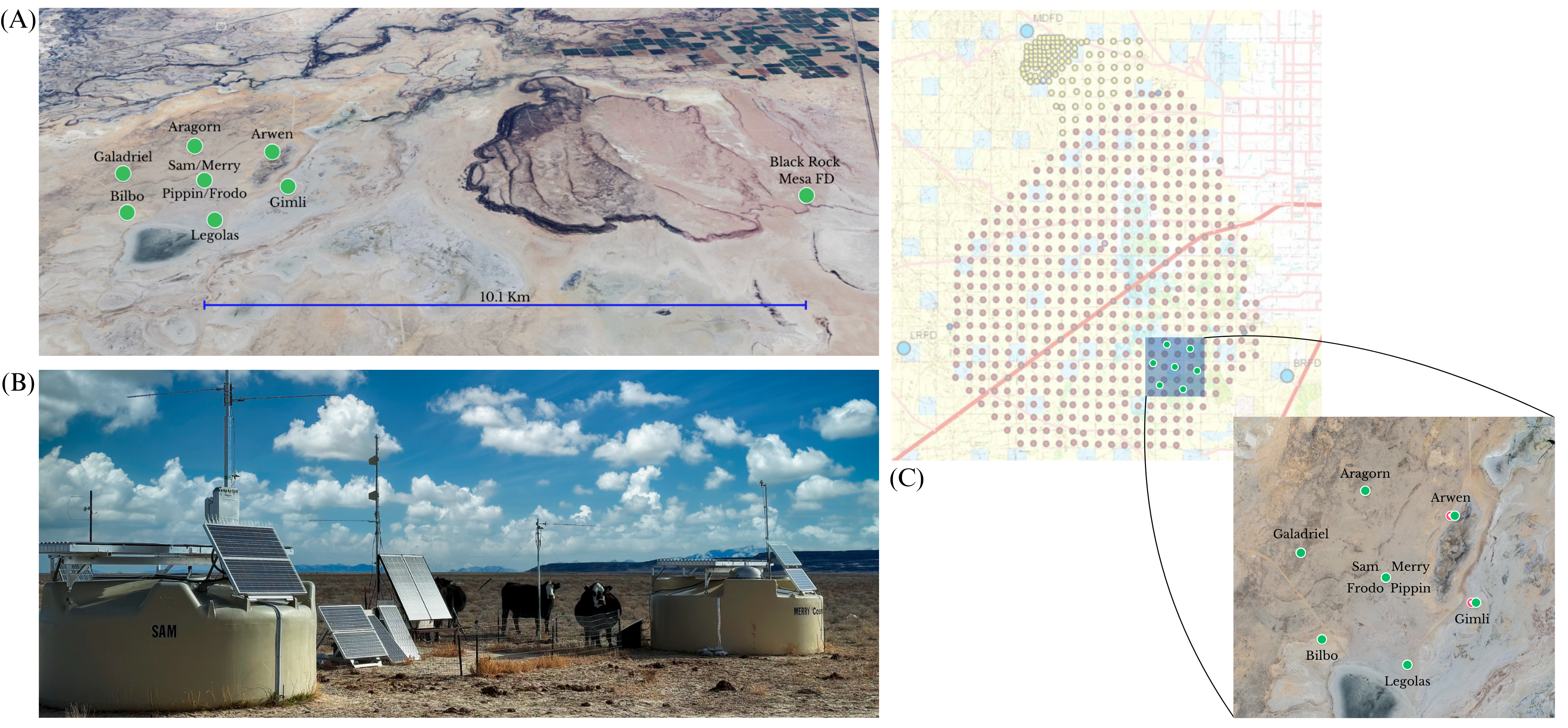}
    \caption{(A) Micro-array location in relation of TA's Black Rock Mesa FD, (B) The Auger@TA central stations, (C) Location of micro-array within TA's site.}
    \label{fig:location}
\end{figure}

% \begin{figure}[ht]
%     \centering
%     \begin{minipage}[t]{0.49\textwidth} % Adjust width as needed
%         \centering
%         \includegraphics[width=\linewidth]{array_loc.png} 
%         \caption{Micro-array location.}
%         \label{fig:location}
%     \end{minipage}
%     \hfill
%     \begin{minipage}[t]{0.49\textwidth}
%         \centering
%         \includegraphics[width=\linewidth]{array_central.jpg}
%         \caption{Central stations.}
%         \label{fig:central}
%     \end{minipage}
% \end{figure}

The main infrastructure of the micro-array comprising the tanks, water, and major structural components were deployed starting in late 2022, southwest of the Telescope Array FD. The deployment area relative to Black Rock Mesa FD, central stations, outrigger antennas, communication base and the location of the micro-array are shown in Figure \ref{fig:location}.

% \begin{wrapfigure}{l}{0.57\linewidth}
%     % \vspace{-2 mm}
%     \centering
%     \includegraphics[width=\linewidth]{augertastation.png}
%     \caption{Auger@TA Station.}
%     \label{fig:augertastation}
% \end{wrapfigure}

\begin{figure}[h]
    \centering
    \includegraphics[width=0.55\textwidth]{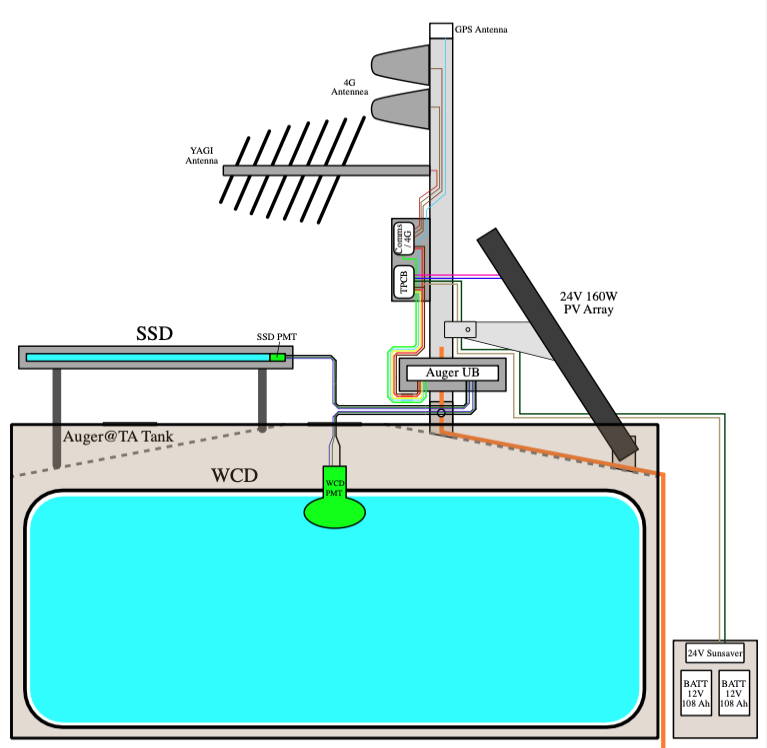}
    \caption{Auger@TA Station.}
    \label{fig:augertastation}
\end{figure}

The stations are designed to closely replicate the specifications of standard Auger stations. A detailed explanation of the modifications, along with a summary of the newly integrated components, can be found in \cite{sonja}, and Figure \ref{fig:augertastation} illustrates the layout of the Auger@TA station. Currently, the key distinctions between an Auger@TA station and an Auger station include the use of a single centrally located PMT, a slightly altered shell design, a custom-built communication system composed of commercially available components, and a PMT board — responsible for distributing and regulating the high-voltage supply to the Photomultiplier Tube (PMT) — handmade for the Auger@TA project due to a slightly different mounting mechanism to the central PMT.

Additionally, all seven stations are equipped with AugerPrime Surface Scintillator Detectors (SSDs), extending the original goal of the deployed pre-upgraded Auger stations and providing enhanced capabilities for cross-calibration.

\section{Communication Systems and Data Handling}

The detector stations communicate with CDAS via radio. For Auger@TA, each station uses a Raspberry Pi 4 for communication purposes and to have a user-friendly system. The Pis are connected to the Unified Board (UB) and integrated with an XBee daughter board connected directly to the Pi to enable radio communication between the station and the central computer where CDAS is run.

The XBee module operates in a 900 MHz band and is paired with an L-COM Yagi antenna at each station. The Yagi antenna is a highly directional device designed to amplify signal strength in a specific direction, and when paired, they significantly enhance the range and reliability of the XBee module, enabling long-range communication. Together, the XBee module and the L-COM Yagi antenna create a highly efficient communication system. 

CDAS is a critical element of Auger, which is responsible for collecting, processing, and storing the data recorded by each detector station. CDAS runs on a Linux-based operating system and is designed as a modular architecture comprising multiple specialized processes communicating through TCP/IP protocols. Each individual detector station runs its own local station data acquisition system, which handles the digitization of signals, triggering, and local storage before transmitting data to CDAS \cite{cdas}.

\begin{figure}[h]
    \centering
    \includegraphics[width=\linewidth]{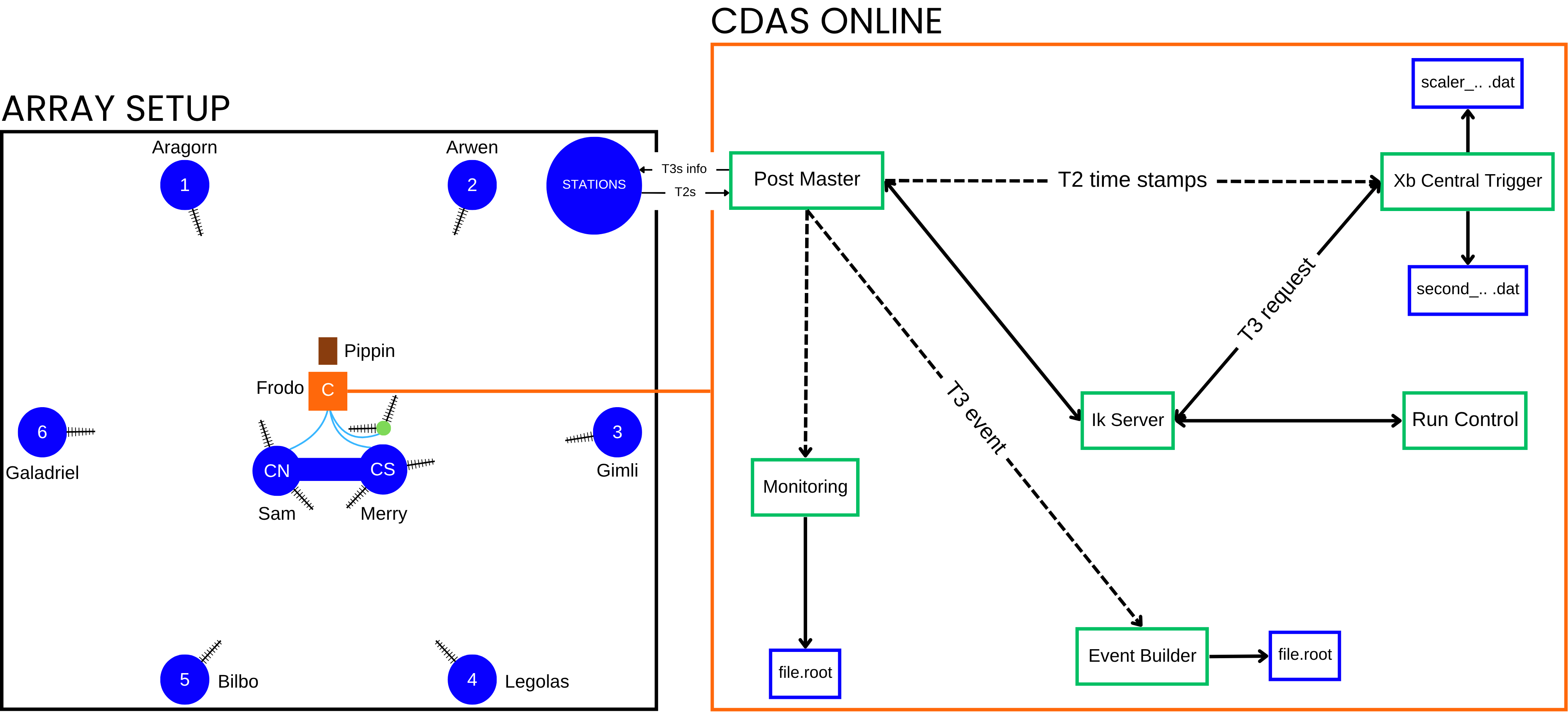}
    \caption{The left side shows the hexagonal micro-array with its nine stations (eight Auger stations and one TA station). Each station communicates with the central computer through a radio antenna installed on a mast on each station. The right side shows CDAS installed and running on Frodo, the central computer and each of the processes and their schematics.}
    \label{fig:diagram}
\end{figure}

Each SD continuously watches for pulses of light in its PMTs. When a pulse exceeds a basic threshold (T1), the station applies a secondary threshold (T2) in which the frequency is set to match the expected rate of cosmic ray signals to the WCD before sending anything to the central system. This two-step verification keeps each station's reported rate from around 100 to around 20 hits per second, ensuring the sent signal has a higher chance of coming from an actual cosmic ray event. All T2 triggers - sent every second - are communicated to the central CDAS, allowing it to look for event-level coincidences across multiple detector stations, checking patterns in time and space that match a real cosmic-ray shower. If three stations report a time-over-threshold (ToT)-T2 within a few milliseconds, the system issues a T3 trigger, and CDAS requests detailed event data from relevant stations. This way, a T3 selects actual shower events, which are recorded and saved in a database at a much lower frequency \cite{auger2010}.

The CDAS software structure involves many key processes. To cite some, there is the Post Master (Pm) that collects and reconstructs incoming fragmented data packets, forwarding them to the appropriate internal clients; the Xb Central Trigger (Xb) searches for event-level coincidences to generate T3 triggers; the Event Builder (Eb) saves the event data from stations; and the IkServer (Ik) acts as an internal routing service, broadcasting messages based on their source and destination specifications. A schematic of the micro-array and CDAS processes can be seen in Figure \ref{fig:diagram}.

In the Auger@TA project, a slightly simplified version of the Auger CDAS was developed, tailored to the smaller scale and specific calibration goals of the micro-array. This version of CDAS employs fewer processes than the main Auger CDAS, reducing system complexity and enhancing reliability in our smaller-scale deployment.

In our adapted CDAS, we replace the standard Auger long-range radio network with an emulator link to talk directly to each UB. This emulator connection allows us to send commands, simulate event-trigger messages, and check data flow and integrity without the complexity and infrastructure of the full Auger radio system.

These adaptations aim to provide a robust yet lightweight DAQ solution for Auger@TA, reliable reconstruction of air showers, and an independent cosmic ray flux measurement. Moving forward, the simplified CDAS framework and emulator capabilities will allow us to rapidly integrate additional stations and efficiently manage data as the full Auger@TA hexagonal micro-array comes online.

\section{Preliminary Observations and Future Prospects}

Following several technical challenges, including humidity-induced cable corrosion, malfunctions in the PMT bases, communication system implementation issues, and the commissioning of certain electronic components, the micro-array is now nearing completion. All stations have been structurally assembled, with only a few PMT bases remaining to be installed, final adjustments to the new base enclosures needed, and integration of the SSDs on top of each tank. 

\begin{figure}[h]
    \centering
    \includegraphics[width=0.8\textwidth]{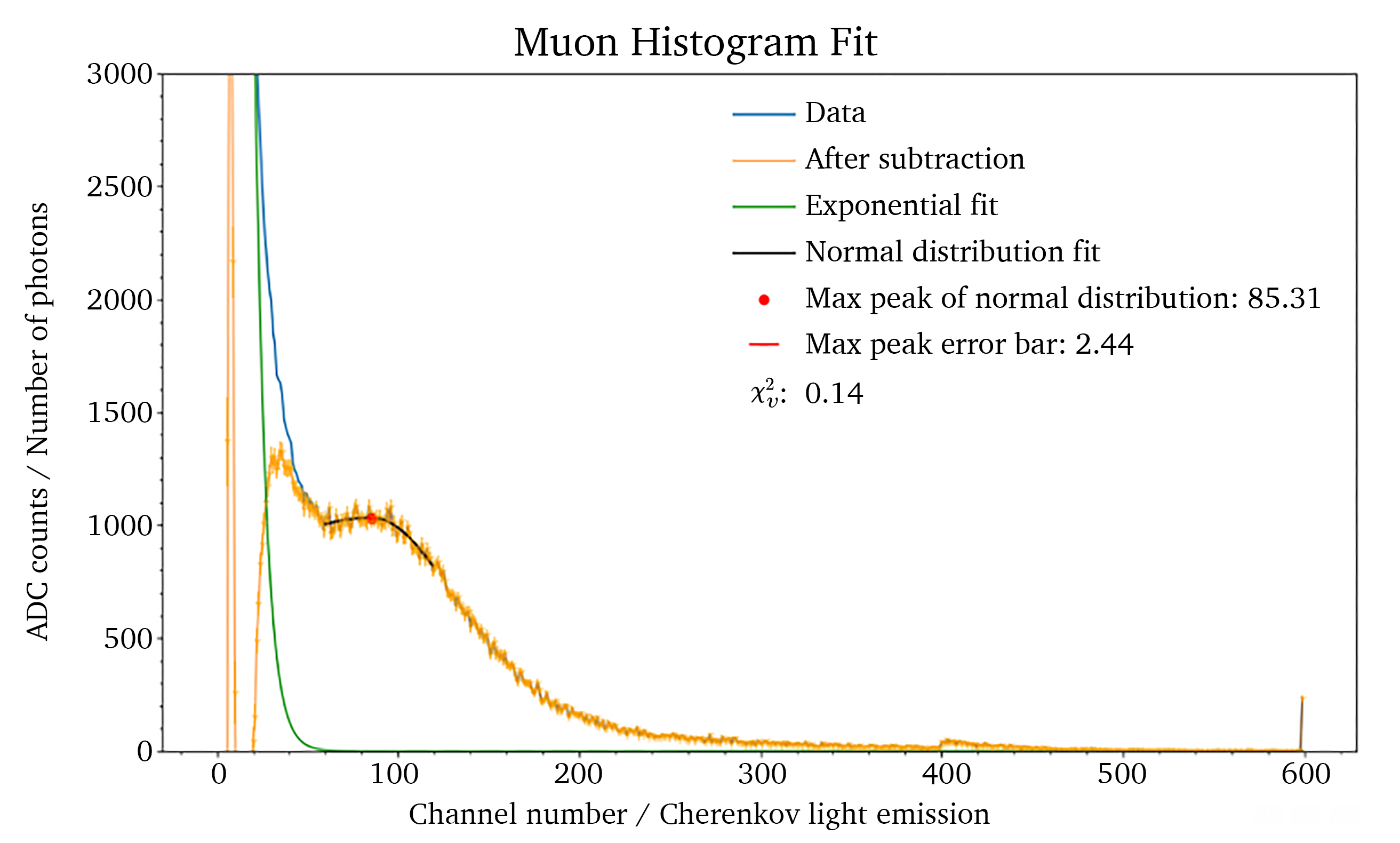}
    \caption{An obtained muon charge histogram for station \textit{Sam}.}
    \label{fig:histogram}
\end{figure}

At present, both Auger stations in the central triplet are fully operational, along with two additional stations that together form a functional triangular sub-array. All installed PMTs are working as expected, passing all tests. An example of a typical charge histogram taken from Sam used for VEM calibration is shown in Figure \ref{fig:histogram}. Raw signal traces have been successfully recorded from these stations, including the central triplet and two outer detectors, along with the corresponding T2 and T3 data. 

\begin{table}[h]
\centering
\begin{tabular}{c|cc|}
\cline{2-3}
\multicolumn{1}{l|}{}              & \multicolumn{2}{c|}{\textbf{Data acquisition (Jan. 2025 - Feb. 2025)}}                          \\ \hline
\multicolumn{1}{|c|}{\textbf{T2s}} & \multicolumn{2}{l|}{20-30 per second per station}                                          \\ \hline
\multicolumn{1}{|c|}{\textbf{T3s}} & \multicolumn{1}{c|}{8 events recorded with 3 stations} & 4 events recorded with 4 stations \\ \hline
\end{tabular}
\caption{Rate of T2 and T3 events recorded during the first test period.}
\label{tab:datatable}
\end{table}

The dedicated communications network explicitly developed for this experiment has proven reliable. As an added precaution, the stations have automatic reboot capabilities if connectivity is lost. All stations except \textit{Legolas} are communicationally operational. The central communication station, \textit{Frodo}, was upgraded in February 2025 with a more robust, watertight enclosure containing a powerful \textit{Intel NUC} mini-PC, that operates the Auger@TA control CDAS.

Data collection is active, for the four operational stations connected to CDAS. As discussed, event detection requires simultaneous triggers from at least three stations arranged in a triangular configuration, such as the existing \textit{Merry/Sam–Galadriel–Bilbo} formation. Thus far, 12 T3 events have been identified, as shown in Table \ref{tab:datatable}.

Efforts are underway to explore methods to analyze coincident signals from the central doublet stations (\textit{Merry/Sam}), which is essential for comparing Auger@TA signals to standard Auger signals. Daily connectivity checks are conducted by exchanging a verification file between the field PC and our data server. Automated scripts monitor this exchange, initiating reboots if communication lapses are detected. Additionally, the CDAS PC generates bi-daily reports detailing temperature, operational uptime fraction, station status, and network connectivity. Lastly, collaboration with the TA team continues at the local TA station, intended for the central triplet configuration. This station is currently being upgraded with new TAx4 electronics. A fully operational micro-array is anticipated by the end of the year.

% REFERENCES %%%%%%%%%%%%%%%%%%%%%%%%%%%%%%%%%%%%%%%%%%%%%%%%%%%%%%%%%%%%%%%%%%%%%%%%%%%%%%%%%%%%%%%%%%%%%%%%%%%%%%%%%%%%%%%%%%%%%%%%%%%%%%%%%%%%%%%%%%%%%%%%%%%%%%%%%%%%%%%%%%%%%%%%

%% Full authors list (ONLY FOR COLLABORATIONS)
\clearpage

\section*{The Pierre Auger Collaboration}

{\footnotesize\setlength{\baselineskip}{10pt}
\noindent
\begin{wrapfigure}[11]{l}{0.12\linewidth}
\vspace{-4pt}
\includegraphics[width=0.98\linewidth]{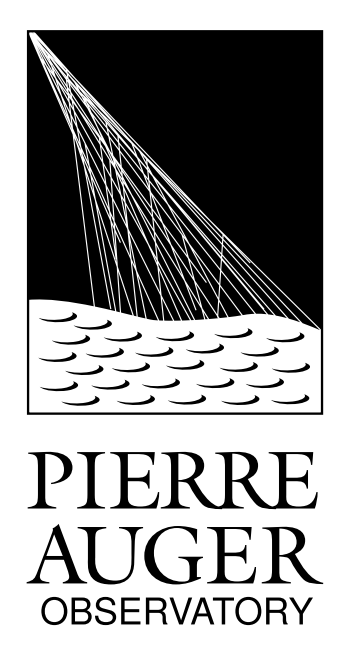}
\end{wrapfigure}
\begin{sloppypar}\noindent
\input{latex_authorlist_authors}
\end{sloppypar}
\begin{center}
\end{center}

\vspace{1ex}
\input{latex_authorlist_institutions}

\input{acknowledgments}
}

\clearpage

\section*{Telescope Array Collaboration}

\begin{wrapfigure}{l}{0.15\linewidth}
    \includegraphics[width=\linewidth]{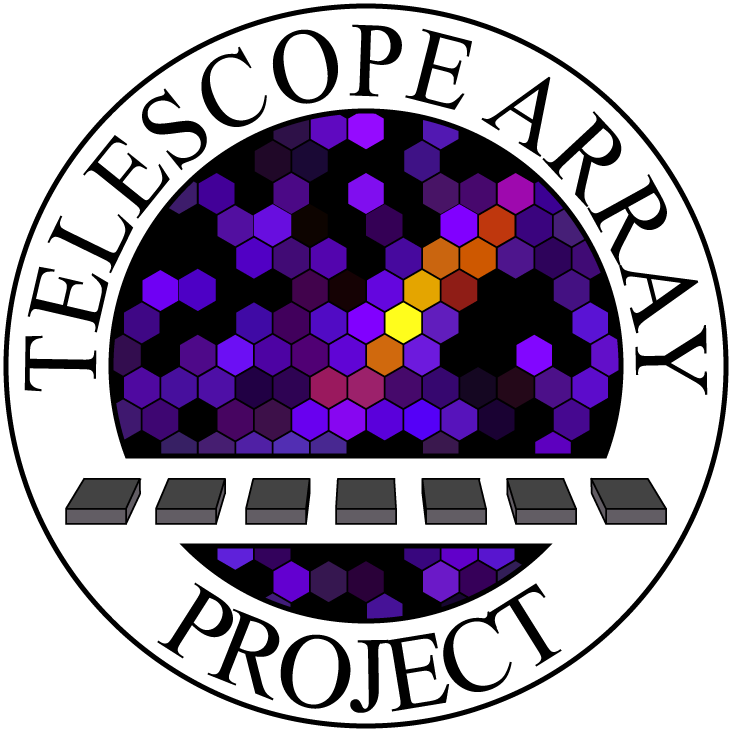}
    \label{fig:auger-ta-spectrum}
\end{wrapfigure}

\noindent R.U. Abbasi$^{1}$, M. Abe$^{2}$, T. Abu-Zayyad$^{1}$, M. Allen$^{1}$, R. Azuma$^{3}$, E. Barcikowski$^{1}$, J.W. Belz$^{1}$, D.R. Bergman$^{1}$, S.A. Blake$^{1}$, R. Cady$^{1}$
% ...

\begin{figure}[h]
    \centering
    \includegraphics[width=0.5\linewidth]{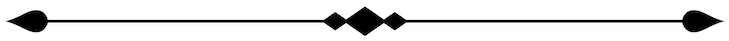}
\end{figure}

\begin{center}
$^{1}$ High Energy Astrophysics Institute and Department of Physics and Astronomy, University of
Utah, Salt Lake City, Utah, USA

$^{2}$ The Graduate School of Science and Engineering, Saitama University, Saitama, Saitama, Japan

$^{3}$ Graduate School of Science and Engineering, Tokyo Institute of Technology, Meguro, Tokyo,
Japan

$^{4}$ Department of Physics and The Research Institute of Natural Science, Hanyang University,
Seongdong-gu, Seoul, Korea

$^{5}$ Department of Physics, Tokyo University of Science, Noda, Chiba, Japan

$^{6}$ Department of Physics, Kinki University, Higashi Osaka, Osaka, Japan

$^{7}$ Institute for Cosmic Ray Research, University of Tokyo, Kashiwa, Chiba, Japan

$^{8}$ Kavli Institute for the Physics and Mathematics of the Universe (WPI), Todai Institutes for
Advanced Study, the University of Tokyo, Kashiwa, Chiba, Japan

$^{9}$ Graduate School of Science, Osaka City University, Osaka, Osaka, Japan

$^{10}$ Information Engineering Graduate School of Science and Technology, Shinshu University,
Nagano, Nagano, Japan

$^{11}$ Faculty of Engineering, Kanagawa University, Yokohama, Kanagawa, Japan

$^{12}$ Interdisciplinary Graduate School of Medicine and Engineering, University of Yamanashi,
Kofu, Yamanashi, Japan

$^{13}$ Astrophysical Big Bang Laboratory, RIKEN, Wako, Saitama, Japan

$^{14}$ Department of Physics, Tokyo City University, Setagaya-ku, Tokyo, Japan

$^{15}$ Institute for Nuclear Research of the Russian Academy of Sciences, Moscow, Russia

$^{16}$ Advanced Research Institute for Science and Engineering, Waseda University, Shinjuku-ku,
Tokyo, Japan

$^{17}$ Department of Physics, Chiba University, Chiba, Chiba, Japan

$^{18}$ Department of Physics, School of Natural Sciences, Ulsan National Institute of Science and
Technology, UNIST-gil, Ulsan, Korea

$^{19}$ Department of Physics, Yonsei University, Seodaemun-gu, Seoul, Korea

$^{20}$ Institute of Particle and Nuclear Studies, KEK, Tsukuba, Ibaraki, Japan

$^{21}$ Faculty of Science, Kochi University, Kochi, Kochi, Japan

$^{22}$ Department of Physical Sciences, Ritsumeikan University, Kusatsu, Shiga, Japan

$^{23}$ Department of Physics, Sungkyunkwan University, Jang-an-gu, Suwon, Korea

$^{24}$ Sternberg Astronomical Institute, Moscow M.V. Lomonosov State University, Moscow, Russia

$^{25}$ Academic Assembly School of Science and Technology Institute of Engineering, Shinshu
University, Nagano, Nagano, Japan

$^{26}$ Department of Physics and Astronomy, Rutgers University - The State University of New
Jersey, Piscataway, New Jersey, USA

$^{27}$ Earthquake Research Institute, University of Tokyo, Bunkyo-ku, Tokyo, Japan

$^{28}$ Graduate School of Information Sciences, Hiroshima City University, Hiroshima, Hiroshima,
Japan

$^{29}$ Service de Physique Th´eorique, Universit´e Libre de Bruxelles, Brussels, Belgium

$^{30}$ National Institute of Radiological Science, Chiba, Chiba, Japan

$^{31}$ National Institute of Chemical Physics and Biophysics, Estonia

$^{32}$ Department of Physics and Institute for the Early Universe, Ewha Womans University,
Seodaaemun-gu, Seoul, Korea

$^{33}$ Department of Physics, Ehime University, Matsuyama, Ehime, Japan
\end{center}

\section*{Acknowledgements}

\footnotesize The Telescope Array experiment is supported by the Japan Society for the Promotion of Science through Grants-in-Aid for Scientific Research on Specially Promoted
Research (21000002) “Extreme Phenomena in the Universe Explored by Highest Energy Cosmic Rays” and for Scientific Research (19104006), and the Inter-University
Research Program of the Institute for Cosmic Ray Research; by the U.S. National
Science Foundation awards PHY-0601915, PHY-1404495, PHY-1404502, and PHY1607727; by the National Research Foundation of Korea (2015R1A2A1A010068
70, 2015R1A2A1A15055344, 2016R1A5A1013277,
2007-0093860, 2016R1A2B4014967); by the Russian Academy of Sciences, RFBR
grant 16-02-00962a (INR), IISN project No. 4.4502.13, and Belgian Science Policy
under IUAP VII/37 (ULB). The foundations of Dr. Ezekiel R. and Edna Wattis
Dumke, Willard L. Eccles, and George S. and Dolores Dor´e Eccles all helped with
generous donations. The State of Utah supported the project through its Economic
Development Board, and the University of Utah through the Office of the Vice President for Research. The experimental site became available through the cooperation
of the Utah School and Institutional Trust Lands Administration (SITLA), U.S.
Bureau of Land Management (BLM), and the U.S. Air Force. We appreciate the
assistance of the State of Utah and Fillmore offices of the BLM in crafting the Plan
of Development for the site. Patrick Shea assisted the collaboration with valuable
advice on a variety of topics. The people and the officials of Millard County, Utah
have been a source of steadfast and warm support for our work which we greatly appreciate. We are indebted to the Millard County Road Department for their efforts
to maintain and clear the roads which get us to our sites. We gratefully acknowledge
the contribution from the technical staffs of our home institutions. An allocation of
computer time from the Center for High Performance Computing at the University
of Utah is gratefully acknowledged.

%\noindent \textbf{Note comment afterwards:} Collaborations have the possibility to provide an authors list in xml format which will be used while generating the DOI entries making the full authors list searchable in databases like Inspire HEP. \\
%
%\scriptsize

%first.author$^1$, 
%second.author$^2$, 
%third.author$^3$ % .... more names
%and 
%last.author$^{n}$ \\
%
%\noindent
%$^1$first.affiliation.
%$^2$second.affiliation. % .... more affiliation
%$^{m}$last.affiliation.

\end{document}

%% file: latex_authorlist_authors.tex
% created on 2025-06-06
A.~Abdul Halim$^{13}$,
P.~Abreu$^{70}$,
M.~Aglietta$^{53,51}$,
I.~Allekotte$^{1}$,
K.~Almeida Cheminant$^{78,77}$,
A.~Almela$^{7,12}$,
R.~Aloisio$^{44,45}$,
J.~Alvarez-Mu\~niz$^{76}$,
A.~Ambrosone$^{44}$,
J.~Ammerman Yebra$^{76}$
% ...

%% file: latex_authorlist_institutions.tex
% created on 2025-06-06
% needs \usepackage{enumitem}
\begin{description}[labelsep=0.2em,align=right,labelwidth=0.7em,labelindent=0em,leftmargin=2em,noitemsep,before={\renewcommand\makelabel[1]{##1 }}]
\item[$^{1}$] Centro At\'omico Bariloche and Instituto Balseiro (CNEA-UNCuyo-CONICET), San Carlos de Bariloche, Argentina
\item[$^{2}$] Departamento de F\'\i{}sica and Departamento de Ciencias de la Atm\'osfera y los Oc\'eanos, FCEyN, Universidad de Buenos Aires and CONICET, Buenos Aires, Argentina
\item[$^{3}$] IFLP, Universidad Nacional de La Plata and CONICET, La Plata, Argentina
\item[$^{4}$] Instituto de Astronom\'\i{}a y F\'\i{}sica del Espacio (IAFE, CONICET-UBA), Buenos Aires, Argentina
\item[$^{5}$] Instituto de F\'\i{}sica de Rosario (IFIR) -- CONICET/U.N.R.\ and Facultad de Ciencias Bioqu\'\i{}micas y Farmac\'euticas U.N.R., Rosario, Argentina
\item[$^{6}$] Instituto de Tecnolog\'\i{}as en Detecci\'on y Astropart\'\i{}culas (CNEA, CONICET, UNSAM), and Universidad Tecnol\'ogica Nacional -- Facultad Regional Mendoza (CONICET/CNEA), Mendoza, Argentina
\item[$^{7}$] Instituto de Tecnolog\'\i{}as en Detecci\'on y Astropart\'\i{}culas (CNEA, CONICET, UNSAM), Buenos Aires, Argentina
\item[$^{8}$] International Center of Advanced Studies and Instituto de Ciencias F\'\i{}sicas, ECyT-UNSAM and CONICET, Campus Miguelete -- San Mart\'\i{}n, Buenos Aires, Argentina
\item[$^{9}$] Laboratorio Atm\'osfera -- Departamento de Investigaciones en L\'aseres y sus Aplicaciones -- UNIDEF (CITEDEF-CONICET), Argentina
\item[$^{10}$] Observatorio Pierre Auger, Malarg\"ue, Argentina
\item[$^{11}$] Observatorio Pierre Auger and Comisi\'on Nacional de Energ\'\i{}a At\'omica, Malarg\"ue, Argentina
\item[$^{12}$] Universidad Tecnol\'ogica Nacional -- Facultad Regional Buenos Aires, Buenos Aires, Argentina
\item[$^{13}$] University of Adelaide, Adelaide, S.A., Australia
\item[$^{14}$] Universit\'e Libre de Bruxelles (ULB), Brussels, Belgium
\item[$^{15}$] Vrije Universiteit Brussels, Brussels, Belgium
\item[$^{16}$] Centro Brasileiro de Pesquisas Fisicas, Rio de Janeiro, RJ, Brazil
\item[$^{17}$] Centro Federal de Educa\c{c}\~ao Tecnol\'ogica Celso Suckow da Fonseca, Petropolis, Brazil
\item[$^{18}$] Instituto Federal de Educa\c{c}\~ao, Ci\^encia e Tecnologia do Rio de Janeiro (IFRJ), Brazil
\item[$^{19}$] Universidade de S\~ao Paulo, Escola de Engenharia de Lorena, Lorena, SP, Brazil
\item[$^{20}$] Universidade de S\~ao Paulo, Instituto de F\'\i{}sica de S\~ao Carlos, S\~ao Carlos, SP, Brazil
\item[$^{21}$] Universidade de S\~ao Paulo, Instituto de F\'\i{}sica, S\~ao Paulo, SP, Brazil
\item[$^{22}$] Universidade Estadual de Campinas (UNICAMP), IFGW, Campinas, SP, Brazil
\item[$^{23}$] Universidade Estadual de Feira de Santana, Feira de Santana, Brazil
\item[$^{24}$] Universidade Federal de Campina Grande, Centro de Ciencias e Tecnologia, Campina Grande, Brazil
\item[$^{25}$] Universidade Federal do ABC, Santo Andr\'e, SP, Brazil
\item[$^{26}$] Universidade Federal do Paran\'a, Setor Palotina, Palotina, Brazil
\item[$^{27}$] Universidade Federal do Rio de Janeiro, Instituto de F\'\i{}sica, Rio de Janeiro, RJ, Brazil
\item[$^{28}$] Universidad de Medell\'\i{}n, Medell\'\i{}n, Colombia
\item[$^{29}$] Universidad Industrial de Santander, Bucaramanga, Colombia
\item[$^{30}$] Charles University, Faculty of Mathematics and Physics, Institute of Particle and Nuclear Physics, Prague, Czech Republic
\item[$^{31}$] Institute of Physics of the Czech Academy of Sciences, Prague, Czech Republic
\item[$^{32}$] Palacky University, Olomouc, Czech Republic
\item[$^{33}$] CNRS/IN2P3, IJCLab, Universit\'e Paris-Saclay, Orsay, France
\item[$^{34}$] Laboratoire de Physique Nucl\'eaire et de Hautes Energies (LPNHE), Sorbonne Universit\'e, Universit\'e de Paris, CNRS-IN2P3, Paris, France
\item[$^{35}$] Univ.\ Grenoble Alpes, CNRS, Grenoble Institute of Engineering Univ.\ Grenoble Alpes, LPSC-IN2P3, 38000 Grenoble, France
\item[$^{36}$] Universit\'e Paris-Saclay, CNRS/IN2P3, IJCLab, Orsay, France
\item[$^{37}$] Bergische Universit\"at Wuppertal, Department of Physics, Wuppertal, Germany
\item[$^{38}$] Karlsruhe Institute of Technology (KIT), Institute for Experimental Particle Physics, Karlsruhe, Germany
\item[$^{39}$] Karlsruhe Institute of Technology (KIT), Institut f\"ur Prozessdatenverarbeitung und Elektronik, Karlsruhe, Germany
\item[$^{40}$] Karlsruhe Institute of Technology (KIT), Institute for Astroparticle Physics, Karlsruhe, Germany
\item[$^{41}$] RWTH Aachen University, III.\ Physikalisches Institut A, Aachen, Germany
\item[$^{42}$] Universit\"at Hamburg, II.\ Institut f\"ur Theoretische Physik, Hamburg, Germany
\item[$^{43}$] Universit\"at Siegen, Department Physik -- Experimentelle Teilchenphysik, Siegen, Germany
\item[$^{44}$] Gran Sasso Science Institute, L'Aquila, Italy
\item[$^{45}$] INFN Laboratori Nazionali del Gran Sasso, Assergi (L'Aquila), Italy
\item[$^{46}$] INFN, Sezione di Catania, Catania, Italy
\item[$^{47}$] INFN, Sezione di Lecce, Lecce, Italy
\item[$^{48}$] INFN, Sezione di Milano, Milano, Italy
\item[$^{49}$] INFN, Sezione di Napoli, Napoli, Italy
\item[$^{50}$] INFN, Sezione di Roma ``Tor Vergata'', Roma, Italy
\item[$^{51}$] INFN, Sezione di Torino, Torino, Italy
\item[$^{52}$] Istituto di Astrofisica Spaziale e Fisica Cosmica di Palermo (INAF), Palermo, Italy
\item[$^{53}$] Osservatorio Astrofisico di Torino (INAF), Torino, Italy
\item[$^{54}$] Politecnico di Milano, Dipartimento di Scienze e Tecnologie Aerospaziali , Milano, Italy
\item[$^{55}$] Universit\`a del Salento, Dipartimento di Matematica e Fisica ``E.\ De Giorgi'', Lecce, Italy
\item[$^{56}$] Universit\`a dell'Aquila, Dipartimento di Scienze Fisiche e Chimiche, L'Aquila, Italy
\item[$^{57}$] Universit\`a di Catania, Dipartimento di Fisica e Astronomia ``Ettore Majorana``, Catania, Italy
\item[$^{58}$] Universit\`a di Milano, Dipartimento di Fisica, Milano, Italy
\item[$^{59}$] Universit\`a di Napoli ``Federico II'', Dipartimento di Fisica ``Ettore Pancini'', Napoli, Italy
\item[$^{60}$] Universit\`a di Palermo, Dipartimento di Fisica e Chimica ''E.\ Segr\`e'', Palermo, Italy
\item[$^{61}$] Universit\`a di Roma ``Tor Vergata'', Dipartimento di Fisica, Roma, Italy
\item[$^{62}$] Universit\`a Torino, Dipartimento di Fisica, Torino, Italy
\item[$^{63}$] Benem\'erita Universidad Aut\'onoma de Puebla, Puebla, M\'exico
\item[$^{64}$] Unidad Profesional Interdisciplinaria en Ingenier\'\i{}a y Tecnolog\'\i{}as Avanzadas del Instituto Polit\'ecnico Nacional (UPIITA-IPN), M\'exico, D.F., M\'exico
\item[$^{65}$] Universidad Aut\'onoma de Chiapas, Tuxtla Guti\'errez, Chiapas, M\'exico
\item[$^{66}$] Universidad Michoacana de San Nicol\'as de Hidalgo, Morelia, Michoac\'an, M\'exico
\item[$^{67}$] Universidad Nacional Aut\'onoma de M\'exico, M\'exico, D.F., M\'exico
\item[$^{68}$] Institute of Nuclear Physics PAN, Krakow, Poland
\item[$^{69}$] University of \L{}\'od\'z, Faculty of High-Energy Astrophysics,\L{}\'od\'z, Poland
\item[$^{70}$] Laborat\'orio de Instrumenta\c{c}\~ao e F\'\i{}sica Experimental de Part\'\i{}culas -- LIP and Instituto Superior T\'ecnico -- IST, Universidade de Lisboa -- UL, Lisboa, Portugal
\item[$^{71}$] ``Horia Hulubei'' National Institute for Physics and Nuclear Engineering, Bucharest-Magurele, Romania
\item[$^{72}$] Institute of Space Science, Bucharest-Magurele, Romania
\item[$^{73}$] Center for Astrophysics and Cosmology (CAC), University of Nova Gorica, Nova Gorica, Slovenia
\item[$^{74}$] Experimental Particle Physics Department, J.\ Stefan Institute, Ljubljana, Slovenia
\item[$^{75}$] Universidad de Granada and C.A.F.P.E., Granada, Spain
\item[$^{76}$] Instituto Galego de F\'\i{}sica de Altas Enerx\'\i{}as (IGFAE), Universidade de Santiago de Compostela, Santiago de Compostela, Spain
\item[$^{77}$] IMAPP, Radboud University Nijmegen, Nijmegen, The Netherlands
\item[$^{78}$] Nationaal Instituut voor Kernfysica en Hoge Energie Fysica (NIKHEF), Science Park, Amsterdam, The Netherlands
\item[$^{79}$] Stichting Astronomisch Onderzoek in Nederland (ASTRON), Dwingeloo, The Netherlands
\item[$^{80}$] Universiteit van Amsterdam, Faculty of Science, Amsterdam, The Netherlands
\item[$^{81}$] Case Western Reserve University, Cleveland, OH, USA
\item[$^{82}$] Colorado School of Mines, Golden, CO, USA
\item[$^{83}$] Department of Physics and Astronomy, Lehman College, City University of New York, Bronx, NY, USA
\item[$^{84}$] Michigan Technological University, Houghton, MI, USA
\item[$^{85}$] New York University, New York, NY, USA
\item[$^{86}$] University of Chicago, Enrico Fermi Institute, Chicago, IL, USA
\item[$^{87}$] University of Delaware, Department of Physics and Astronomy, Bartol Research Institute, Newark, DE, USA
\item[] -----
\item[$^{a}$] Max-Planck-Institut f\"ur Radioastronomie, Bonn, Germany
\item[$^{b}$] also at Kapteyn Institute, University of Groningen, Groningen, The Netherlands
\item[$^{c}$] School of Physics and Astronomy, University of Leeds, Leeds, United Kingdom
\item[$^{d}$] Fermi National Accelerator Laboratory, Fermilab, Batavia, IL, USA
\item[$^{e}$] Pennsylvania State University, University Park, PA, USA
\item[$^{f}$] Colorado State University, Fort Collins, CO, USA
\item[$^{g}$] Louisiana State University, Baton Rouge, LA, USA
\item[$^{h}$] now at Graduate School of Science, Osaka Metropolitan University, Osaka, Japan
\item[$^{i}$] Institut universitaire de France (IUF), France
\item[$^{j}$] now at Technische Universit\"at Dortmund and Ruhr-Universit\"at Bochum, Dortmund and Bochum, Germany
\end{description}

%% file: acknowledgments.tex
% created on 2025-06-06
\section*{Acknowledgments}

\begin{sloppypar}
The successful installation, commissioning, and operation of the Pierre
Auger Observatory would not have been possible without the strong
commitment and effort from the technical and administrative staff in
Malarg\"ue. We are very grateful to the following agencies and
organizations for financial support:
\end{sloppypar}

\begin{sloppypar}
Argentina -- Comisi\'on Nacional de Energ\'\i{}a At\'omica; Agencia Nacional de
Promoci\'on Cient\'\i{}fica y Tecnol\'ogica (ANPCyT); Consejo Nacional de
Investigaciones Cient\'\i{}ficas y T\'ecnicas (CONICET); Gobierno de la
Provincia de Mendoza; Municipalidad de Malarg\"ue; NDM Holdings and Valle
Las Le\~nas; in gratitude for their continuing cooperation over land
access; Australia -- the Australian Research Council; Belgium -- Fonds
de la Recherche Scientifique (FNRS); Research Foundation Flanders (FWO),
Marie Curie Action of the European Union Grant No.~101107047; Brazil --
Conselho Nacional de Desenvolvimento Cient\'\i{}fico e Tecnol\'ogico (CNPq);
Financiadora de Estudos e Projetos (FINEP); Funda\c{c}\~ao de Amparo \`a
Pesquisa do Estado de Rio de Janeiro (FAPERJ); S\~ao Paulo Research
Foundation (FAPESP) Grants No.~2019/10151-2, No.~2010/07359-6 and
No.~1999/05404-3; Minist\'erio da Ci\^encia, Tecnologia, Inova\c{c}\~oes e
Comunica\c{c}\~oes (MCTIC); Czech Republic -- GACR 24-13049S, CAS LQ100102401,
MEYS LM2023032, CZ.02.1.01/0.0/0.0/16{\textunderscore}013/0001402,
CZ.02.1.01/0.0/0.0/18{\textunderscore}046/0016010 and
CZ.02.1.01/0.0/0.0/17{\textunderscore}049/0008422 and CZ.02.01.01/00/22{\textunderscore}008/0004632;
France -- Centre de Calcul IN2P3/CNRS; Centre National de la Recherche
Scientifique (CNRS); Conseil R\'egional Ile-de-France; D\'epartement
Physique Nucl\'eaire et Corpusculaire (PNC-IN2P3/CNRS); D\'epartement
Sciences de l'Univers (SDU-INSU/CNRS); Institut Lagrange de Paris (ILP)
Grant No.~LABEX ANR-10-LABX-63 within the Investissements d'Avenir
Programme Grant No.~ANR-11-IDEX-0004-02; Germany -- Bundesministerium
f\"ur Bildung und Forschung (BMBF); Deutsche Forschungsgemeinschaft (DFG);
Finanzministerium Baden-W\"urttemberg; Helmholtz Alliance for
Astroparticle Physics (HAP); Helmholtz-Gemeinschaft Deutscher
Forschungszentren (HGF); Ministerium f\"ur Kultur und Wissenschaft des
Landes Nordrhein-Westfalen; Ministerium f\"ur Wissenschaft, Forschung und
Kunst des Landes Baden-W\"urttemberg; Italy -- Istituto Nazionale di
Fisica Nucleare (INFN); Istituto Nazionale di Astrofisica (INAF);
Ministero dell'Universit\`a e della Ricerca (MUR); CETEMPS Center of
Excellence; Ministero degli Affari Esteri (MAE), ICSC Centro Nazionale
di Ricerca in High Performance Computing, Big Data and Quantum
Computing, funded by European Union NextGenerationEU, reference code
CN{\textunderscore}00000013; M\'exico -- Consejo Nacional de Ciencia y Tecnolog\'\i{}a
(CONACYT) No.~167733; Universidad Nacional Aut\'onoma de M\'exico (UNAM);
PAPIIT DGAPA-UNAM; The Netherlands -- Ministry of Education, Culture and
Science; Netherlands Organisation for Scientific Research (NWO); Dutch
national e-infrastructure with the support of SURF Cooperative; Poland
-- Ministry of Education and Science, grants No.~DIR/WK/2018/11 and
2022/WK/12; National Science Centre, grants No.~2016/22/M/ST9/00198,
2016/23/B/ST9/01635, 2020/39/B/ST9/01398, and 2022/45/B/ST9/02163;
Portugal -- Portuguese national funds and FEDER funds within Programa
Operacional Factores de Competitividade through Funda\c{c}\~ao para a Ci\^encia
e a Tecnologia (COMPETE); Romania -- Ministry of Research, Innovation
and Digitization, CNCS-UEFISCDI, contract no.~30N/2023 under Romanian
National Core Program LAPLAS VII, grant no.~PN 23 21 01 02 and project
number PN-III-P1-1.1-TE-2021-0924/TE57/2022, within PNCDI III; Slovenia
-- Slovenian Research Agency, grants P1-0031, P1-0385, I0-0033, N1-0111;
Spain -- Ministerio de Ciencia e Innovaci\'on/Agencia Estatal de
Investigaci\'on (PID2019-105544GB-I00, PID2022-140510NB-I00 and
RYC2019-027017-I), Xunta de Galicia (CIGUS Network of Research Centers,
Consolidaci\'on 2021 GRC GI-2033, ED431C-2021/22 and ED431F-2022/15),
Junta de Andaluc\'\i{}a (SOMM17/6104/UGR and P18-FR-4314), and the European
Union (Marie Sklodowska-Curie 101065027 and ERDF); USA -- Department of
Energy, Contracts No.~DE-AC02-07CH11359, No.~DE-FR02-04ER41300,
No.~DE-FG02-99ER41107 and No.~DE-SC0011689; National Science Foundation,
Grant No.~0450696, and NSF-2013199; The Grainger Foundation; Marie
Curie-IRSES/EPLANET; European Particle Physics Latin American Network;
and UNESCO.
\end{sloppypar}